\documentclass[hidelinks]{aa}  

\usepackage{graphicx}
\usepackage{txfonts}
\usepackage{lipsum}
\usepackage{subcaption}
\usepackage{hyperref}
\usepackage{natbib}
\usepackage{tikz}
\usepackage{comment}

\usetikzlibrary{shapes.geometric, arrows, positioning}

\definecolor{darkyellow}{RGB}{254,153,0}

\tikzstyle{box} = [rectangle, rounded corners, minimum width=5cm, minimum height=1.5cm, text centered, draw=black, fill=yellow!60, text=black]
\tikzstyle{smallbox} = [rectangle, rounded corners, minimum width=5cm, minimum height=1cm, text centered, draw=black, fill=yellow!40, text=black]
\tikzstyle{arrow} = [thick,->,>=stealth, black]

\usepackage{lscape}            
\usepackage{placeins}           
\usepackage{CJKutf8}

\usepackage{bm}
\usepackage{ulem}           
\usepackage{amsmath}

\begin{document}

   \title{Self-induced edge rings in protoplanetary disks}

   \author{
   Massimiliano Bolchini\inst{1}\fnmsep\inst{2}\fnmsep\inst{3}\fnmsep\inst{4}
   \and
   Haochang Jiang \begin{CJK*}{UTF8}{gbsn}(蒋昊昌)\end{CJK*}\inst{2}
   \and
   Jiaqing Bi \begin{CJK*}{UTF8}{gbsn}(毕嘉擎)\end{CJK*}\inst{2}}

   \institute{
   Università degli Studi di Milano, Via Giovanni Celoria 16, I-20133 Milano, Italy
   \and 
   Max-Planck-Institut für Astronomie (MPIA), Königstuhl 17, 69117 Heidelberg, Germany
   \and
   Dipartimento di Fisica e Astronomia, Università di Bologna, Via Gobetti 93/2, 40129 Bologna, Italy
   \and
   INAF – Osservatorio di Astrofisica e Scienza dello Spazio di Bologna, Via Gobetti 93/3, 40129 Bologna, Italy\\
   \email{massimilian.bolchin2@unibo.it}
   }

   \date{Received September 30, 20XX}

  \abstract
   {Observations with a high angular resolution by ALMA have revealed that substructures are ubiquitous in protoplanetary disks. Axisymmetric dust rings are the most common morphology. The profiles of some observed disks, including young disks, are smooth overall, with a localized dip or bump near the outer edge of the continuum disk that manifests as an edge ring. While embedded planets might contribute to these structures, their physical origin remains unclear.}
   {We investigated the possibility that these edge rings arise purely from radiative transfer effects at the outer edge of a protoplanetary disk. A steep surface density dust gradient at the outer edge of the disk allows stellar irradiation to penetrate more efficiently beyond the disk edge, producing a non-monotonic temperature profile characterized by a dip that is followed by a bump. We tested whether this non-monotonic temperature structure can generate and maintain a localized continuum enhancement.} 
   {We coupled radiative transfer and dust evolution by iterating between the Monte Carlo radiative transfer code RADMC-3D and the dust evolution code DustPy. This framework self-consistently follows the coupled evolution of temperature, grain growth, and dust dynamics.}
   {Thermodynamic feedback at the disk edge can naturally generate and maintain localized dust enhancements resembling the edge rings that are observed in some extremely young disks. Without invoking planets or additional dynamical perturbations, this mechanism offers a plausible explanation for the first-generation ring formation.}
   {Our results highlight the importance of coupling thermodynamics and dust evolution when modeling protoplanetary disks, suggesting that thermodynamic feedback probably plays a role in shaping disk substructures.}

   \keywords{Protoplanetary disks --
                Radiative transfer 
               }

   \maketitle

\nolinenumbers

\section{Introduction} \label{sec: intro}

In recent years, the Atacama Large Millimeter/submillimeter Array (ALMA) has revolutionized our view of planet-forming disks through its unprecedented angular resolution and sensitivity \citep{andrews.SM.2018.12, long.F.2018.12, cieza.LA.2019.01, oberg.KI.2021.11, teague.R.2025.04}. Among these observations, continuum emission traces the thermal emission from millimeter- and submillimeter-sized dust grains in protoplanetary disks, revealing that the spatial distribution of these grains is not as smooth as previously expected in the solar nebula model \citep{weidenschilling.SJ.1977.09}, but is instead highly structured \citep{bae.J.2023.07}. These disk substructures are ubiquitous and appear in a variety of forms, including spirals (\citealt{perez.LM.2016.09, dong.R.2018.06, huang.J.2018.12a, rosotti.GP.2020.06, yoshida.TC.2025.09}), arcs (\citealt{fukagawa.M.2013.12, vandermarel.N.2013.06, isella.A.2018.12, wolfer.L.2025.04}), and, most commonly, annular rings (\citealt{almapartnership.CL.2015.07, andrews.SM.2016.04, isella.A.2016.12, perez.S.2019.07, bosschaart.Q.2026.02}).

Dust rings are of particular interest because they imply a local accumulation of dust grains and are thought to offer ideal conditions for planet formation \citep{pinilla.P.2012.02, jiang.H.2023.01}. In the core-accretion scenario \citep{pollack.JB.1996.11}, planet formation requires sub-micron-sized dust grains inherited from the interstellar medium \citep{Mathis1977} to grow from sub-micron grains to planetary bodies through mutual collisions. This growth must occur within the disk lifetime of only $\sim$1–10 Myr \citep{Haisch2001, mamajek.EE.2009.08}, which makes efficient dust concentration a key ingredient of planet formation.

Several mechanisms have been proposed to explain the formation of dust rings. One proposed mechanism involves ice lines, which mark the radial locations where volatile species condense from the gas phase onto dust grains. Near these transitions, changes in grain composition and sticking properties can modify the efficiency of grain growth and fragmentation, potentially leading to local dust enhancements and ring formation \citep{Zhang_iceline, Okuzumi_sintering, Pinilla_iceline}. Another possibility is the formation of pressure bumps, for instance, via magnetohydrodynamical effects \citep{Flock_2015, pinilla_MHD, Suriano_2017, Suriano_2018}. Such pressure bumps can efficiently trap dust particles, leading to the formation of a dust ring \citep{Pinilla_review_2025}.

Among the proposed mechanisms, gravitational interactions between planets and the disk are one of the most commonly invoked explanations for ring formation. A sufficiently massive planet can open a gap in the gas disk, creating a pressure maximum at the outer edge of the gap \citep{Lin_Papaloizou, Crida2006, Dipierro2016, Rosotti2016}.
However, attributing all observed rings to planets raises a chicken-and-egg problem: if rings are exclusively the outcome of planet formation, it becomes unclear how the first planet could form in the absence of a mechanism that concentrates dust beforehand. This has motivated the exploration of alternative processes capable of generating dust rings without requiring preexisting planets \citep[e.g.,][]{tominaga.RT.2020.09,ohashi.S.2021.02,Jiang_2021}. 

The planet-free scenario is particularly relevant to a peculiar subset of disks: very young systems in which planets are not yet expected to have formed, but dust rings have already been observed. In these objects, the radial profile of the continuum emission is typically smooth with a single dip or bump at the outer disk edge, resembling a pizza crust, for example, around the intermediate T~Tauri star CR~Cha \citep[][]{CrCha} and the Class 0 Oph~A~SM1 \citep[][]{OphSM1}. In both cases and especially in the latter, the young age prevents the majority of the mechanisms, which require timescales that are too long to develop the ring. For CR~Cha, \cite{kim.S.2020.01} suggested that such edge rings might alternatively arise from radiative transfer effects at the disk edge. In this scenario, they proposed that a pressure bump can form from a non-monotonic temperature structure in disks with a steep surface density dust gradient at the outer edge. This temperature structure is produced by two effects: beyond the dust disk edge, the optical depth to stellar irradiation drops sharply, allowing more stellar radiation to heat the optically thin midplane. On the other hand, large grains, which contribute efficiently to radiative cooling, are absent from this region because the radial drift is more efficient, thereby reducing the local cooling efficiency. However, this idea is qualitative and remains quantitatively untested.

We revisit this problem using a coupled modeling framework that self-consistently evolves the thermal disk structure and the dust distribution. Unlike previous qualitative studies, our model explicitly accounts for the feedback between radiative transfer and dust evolution. We show that temperature variations induced at the disk edge can naturally lead to the formation and maintenance of a dust ring. Our results demonstrate that thermodynamic feedback associated with radiative transfer can generate and sustain edge rings, providing an alternative explanation for these structures that does not require embedded planets or additional dynamical mechanisms.

The plan for the paper is as follows: In Sect.~\ref{sec: methods} we describe the hydrodynamical setup and dust coagulation model (Sect.~\ref{sec: hydro}), the radiative transfer setup (Sect.~\ref{sec: RT}), and the iteration procedure (Sect.~\ref{sec: iteration}). The results are presented in Sect.~\ref{sec: results}, and we discuss their implications and conclude in Sect.~\ref{sec: conclusions}.

\section{Methods}
\label{sec: methods}
We considered a viscous protoplanetary disk model of mass $M_{\rm disk}$ orbiting a central star of mass $M_\odot$. The disk consisted of the gas and dust components, with the dust population represented by seven species spanning grain sizes from 1 nm to 1 mm.

\subsection{Hydrodynamics and grain size evolution}
\label{sec: hydro}

We first considered a 1D axisymmetric disk model in the midplane. The model spanned 5 to 200 au in the radial direction, with 128 logarithmically spaced cells. The time evolution of gas and dust was computed using the code \texttt{DustPy} \citep{Dustpy}, and we mainly followed the standard model provided by the code.

Our fiducial total mass of the disk was $M_{\rm disk} = 0.1 M_\odot$, which was inspired by recent observational estimates of massive young disks \citep[e.g.,][]{exoalma_trapman}. We adopted an isothermal equation of state $P = \rho_{\rm g}c_{\rm s}^2$, where $\{P,\rho_{\rm g},c_{\rm s}\}$ are the pressure, midplane density, and sound speed of the gas, respectively. The vertically isothermal temperature of the gas, $T_{\rm g}$, was set equal to the temperature of the smallest dust species, which was to be calculated via radiative transfer simulations (see Sect.~\ref{sec: RT}). The temperature of the gas determines the sound speed via
\begin{equation}
    c_{\rm s} = \sqrt{\frac{k_{\rm B}T_{\rm g}}{\mu m_{\rm u}}},
\end{equation}
where $k_{\rm B}$ is the Boltzmann constant, $\mu = 2.34$ is the mean molecular weight, and $m_{\rm u}$ is the atomic mass unit. 

The hydrodynamic equation for the gas is given by 
\begin{equation}
   \frac{\partial \Sigma_{\rm g}}{\partial t} + \frac{1}{r}\frac{\partial}{\partial r}\left(r\Sigma_{\rm g}v_{\rm g}\right) = 0,
\end{equation}
where $r$ is the distance to the central star, and $\{\Sigma_{\rm g}, v_{\rm g}\}$ are the surface density and radial velocity of the gas, respectively. The latter was always set to the viscous velocity \citep{lynden-bell.D.1974.09},
\begin{equation}
    v_{\rm g} = -\frac{3}{\Sigma_{\rm g}r^{1/2}}\frac{\partial}{\partial r}\left(\nu \Sigma_{\rm g} r^{1/2}\right).
\end{equation}
The kinematic viscosity $\nu$ is given by $\nu = \alpha c_{\rm s}H_{\rm g} \label{eq: viscosity}$, where $\alpha$ is the Shakura-Sunyaev viscosity parameter \citep{Shakura_Sunyaev_prescription}, $H_{\rm g} = c_{\rm s}/\Omega_{\rm K}$ is the gas scale height, and $\Omega_{\rm K}$ is the Keplerian angular velocity. For all of our models, we adopted a constant $\alpha$ prescription with $\alpha = 10^{-3}$, which is consistent with but toward the higher end of the range inferred from observations \citep{Villenave_2020, Rosotti_2023}.

The initial gas surface density profile is given by
\begin{equation}
    \Sigma_{\rm g} = \Sigma_{\rm g0}\left(\frac{r}{r_{0, \rm g}}\right)^{-p_g}\exp\left[-\left(\frac{r}{r_{\rm c, \rm g}}\right)^{n_g}\right],
    \label{eq: density_profile}
\end{equation}
where $\Sigma_{\rm g0}$ is a normalization factor, $p_g = 1$, $r_{0,\rm g} = 10$ au, $r_{\rm c, \rm g} = 80$ au, and $n_g = 1$.

Dust in our models was modeled as a pressure-less fluid. The dust population was divided into seven logarithmically spaced mass bins (i.e., seven dust species), each characterized by a representative grain size $a$. We adopted $a \in \{10^{-3}, 10^{-2}, \ldots, 10^2, 10^3\}$ microns and assumed a uniform internal grain density of $\rho_{\rm s} = 1.67$ g/cm$^3$. The corresponding representative mass of each bin was then $m = (4/3)\pi\rho_{\rm s} a^3$, and each bin spanned the mass range from $2m/1001$ to $2000m/1001$.

The dust fluids evolve under gas drag, radial drift, and turbulent diffusion, and we neglected the dust back-reaction onto the gas. The hydrodynamic equation for each dust species is given by
\begin{equation}
   \frac{\partial \Sigma_{\rm d}}{\partial t} + \frac{1}{r}\frac{\partial}{\partial r}\left(r\Sigma_{\rm d}v_{\rm d}\right) - \frac{1}{r}\frac{\partial}{\partial r}\left[rD\Sigma_{\rm g}\frac{\partial}{\partial r}\left(\frac{\Sigma_{\rm d}}{\Sigma_{\rm g}}\right)\right] = 0,
\end{equation}
where $\Sigma_{\rm d}$ is the surface density, $v_{\rm d}$ is the radial velocity, and $D$ is the radial diffusion coefficient of each dust species, respectively.
The initial dust surface density profile is 
\begin{equation}
    \Sigma_{\rm d}=\Sigma_{\rm d0} \left(\frac{r}{r_{0}}\right)^{-p}\exp\left[-\left(\frac{r}{r_{\rm c}}\right)^n\right],
\end{equation}
where $p = 1$, $r_0 = 10$ au,$r_{\rm c} = 40$ au, and $n = 10$.
We note that a sufficiently high value of $n$ is needed to produce the steep outer disk edge for the mechanism described in Sect.~\ref{sec: intro}. We also show a comparison of the midplane gas temperature with different $n$ in Fig.~\ref{fig:bump_high_n}.

The normalization of the dust surface density profile $\Sigma_{\rm d0}$ was initialized by adopting a total dust-to-gas mass ratio of 0.01, with the mass in each dust species distributed according to the grain size distribution $n(a) \propto a^{-3.5}$ \citep{Mathis1977}.

The dynamic coupling between gas and dust is parametrized by the Stokes number in the Epstein regime \citep{Epstein},
\begin{equation}
    {\rm St} = \frac{\pi\rho_{\rm s} a}{2\Sigma_{\rm g}}.
\end{equation}
The radial dust velocity is then set to
\begin{equation}
    v_{\rm d} = \frac{v_{\rm g} - 2r\eta\Omega_{\rm K}{\rm St}}{1 + {\rm St}^2},
\end{equation}
where the midplane pressure gradient parameter $\eta$ is
\begin{equation}
    \eta = -\frac{1}{2}\left(\frac{H_{\rm g}}{r}\right)^2\frac{\partial\log P}{\partial\log r}.
    \label{eq: eta}
\end{equation}
The radial dust diffusion coefficient is given by \citep{Youdin_Lithwick_2007}
\begin{equation}
    D = \frac{\delta_{\rm r} c_{\rm s}^2}{\Omega_{\rm K}(1 + {\rm St}^2)},
    \label{eq: dust_diffusion}
\end{equation}
where $\delta_{\rm r}$ is analogous to the gas viscosity parameter $\alpha$, but was treated here as an independent parameter. This allowed us to modify the dust diffusion level in the following sections without affecting the gas dynamics.

The size evolution of dust was computed by solving the Smoluchowski equation with fragmentation. The maximum grain size was set by the fragmentation and drift barriers, following the prescriptions in \citet{birnstiel.T.2012.03}. Our fiducial fragmentation velocity was $v_{\rm frag} = 1$ m/s, which is consistent with laboratory experiments and observational constraints \citep{Haochang2024}. The collision rate in our models depends on the relative velocity between dust grains. This relative velocity is affected, although not exclusively, by turbulence-induced motions, whose strength is controlled by the parameter $\delta_{\rm t}$ with a fiducial value $\delta_{\rm t}= 10^{-3}$. We further note that $\delta_{\rm t}$, $\delta_{\rm r}$, and $\alpha$ were treated as mutually independent parameters in our models. As a result, variation in $\delta_{\rm t}$ only affected the collisional velocity and the collision rate, but did not alter the level of dust diffusion or the viscous gas evolution. We carried out a parameter study to examine the roles of $v_{\rm frag}$, $\delta_{\rm t}$, $\delta_{\rm r}$, and $M_{\rm disk}$; the corresponding model setups are summarized in Table~\ref{table:1}.

\subsection{Radiative transfer}
\label{sec: RT}

The dust temperatures in our models were calculated using the Monte-Carlo-based radiative transfer code \texttt{RADMC-3D} \citep{dullemond.CP.2012.02} with $10^8$ photon packages. The 1D disk model described in Sect.~\ref{sec: hydro} was extended to 2D by incorporating the vertical direction. The radiative transfer model adopted the same radial setup as the hydrodynamical model, and it was azimuthally symmetric, so that the grid in the $\phi$ direction had only one cell.
We assumed that the gas was vertically isothermal and that gas and dust were in vertical hydrostatic equilibrium. Under these assumptions, the scale height of each dust species is given by
\begin{equation}
    H_{\rm d} = \left(1 + \frac{\rm St}{\alpha}\right)^{-1/2}H_{\rm g}.
    \label{eq: scale_height}
\end{equation}
By further assuming that the gas and the smallest dust species shared the same temperature in the midplane, we iteratively solved for the dust temperature until a self-consistent solution for gas temperature and gas scale height was obtained. Because convergence was typically achieved within four iterations, we adopted four iterations as our standard procedure (see Fig.~\ref{fig: code_workflow}).

We considered a disk model surrounding a Sun-like star with a blackbody temperature of $T_\star = T_{\odot}$. The radiative transfer model considered all dust species in the hydrodynamical model. The scattering was treated as isotropic and a wavelength grid from $0.1\,\mu\mathrm{m}$ to $1\,\mathrm{cm}$ was adopted, using piecewise logarithmic sampling with 20, 100, and 30 points in the intervals $0.1$--$7\,\mu\mathrm{m}$, $7$--$25\,\mu\mathrm{m}$, and $25\,\mu\mathrm{m}$--$1\,\mathrm{cm}$, respectively, to provide a higher resolution in the mid-infrared, where dust emission features are most prominent, while keeping a coarser sampling at shorter and longer wavelengths, where the spectrum varies more smoothly.
The dust opacity at each wavelength point was derived using \texttt{Optool} \citep{optool}, assuming that all dust grains are composed of $70\%$ pyroxene and $30\%$ amorphous carbon.

\subsection{Iteration between models}
\label{sec: iteration}
\begin{figure}[h!]
   \centering
   \includegraphics[width=\hsize]{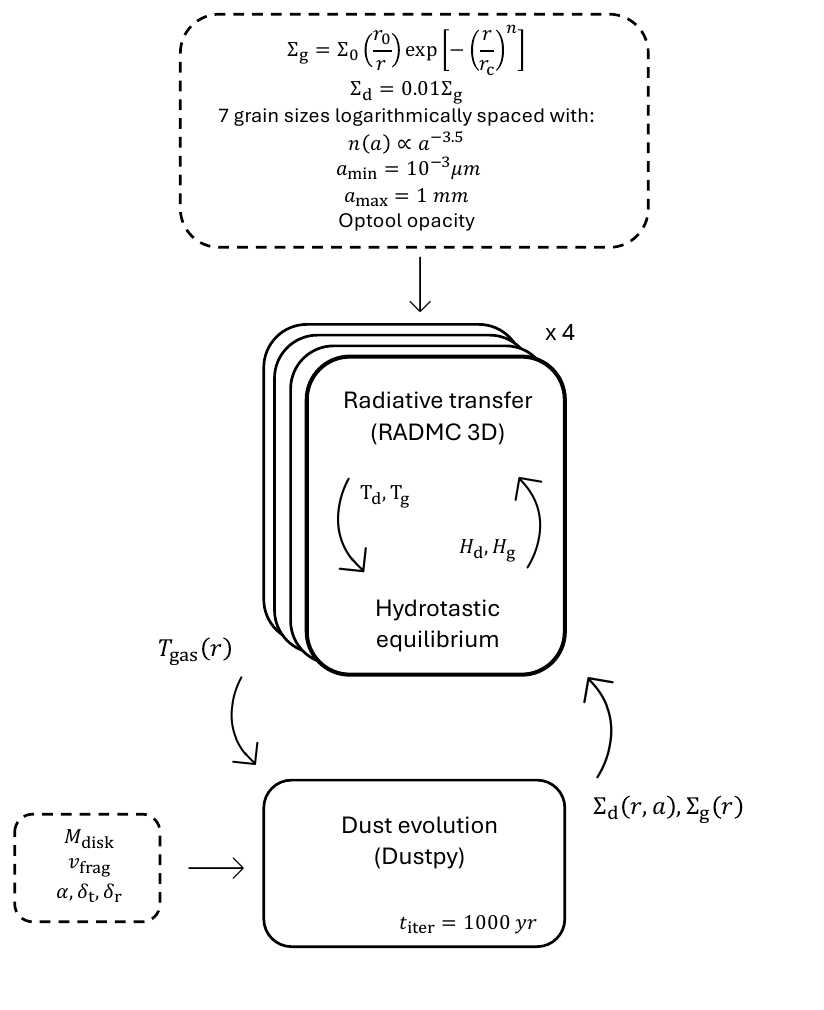}
      \caption{Workflow of the coupled dust evolution and radiative transfer modeling.
      }
         \label{fig: code_workflow}
   \end{figure}

The gas temperature affects the pressure profile, thereby regulating dust dynamics and the efficiency of grain growth. Conversely, dust dynamics and grain growth affect gas temperature structure by modifying the optical depth profile of the disk. These processes are therefore strongly coupled and should be treated self-consistently. To this end, we iteratively coupled our \texttt{DustPy} and \texttt{RADMC-3D} models, following the procedure described below and illustrated in Fig.~\ref{fig: code_workflow}.

\begin{figure}
         \centering
         \includegraphics[width=\hsize]{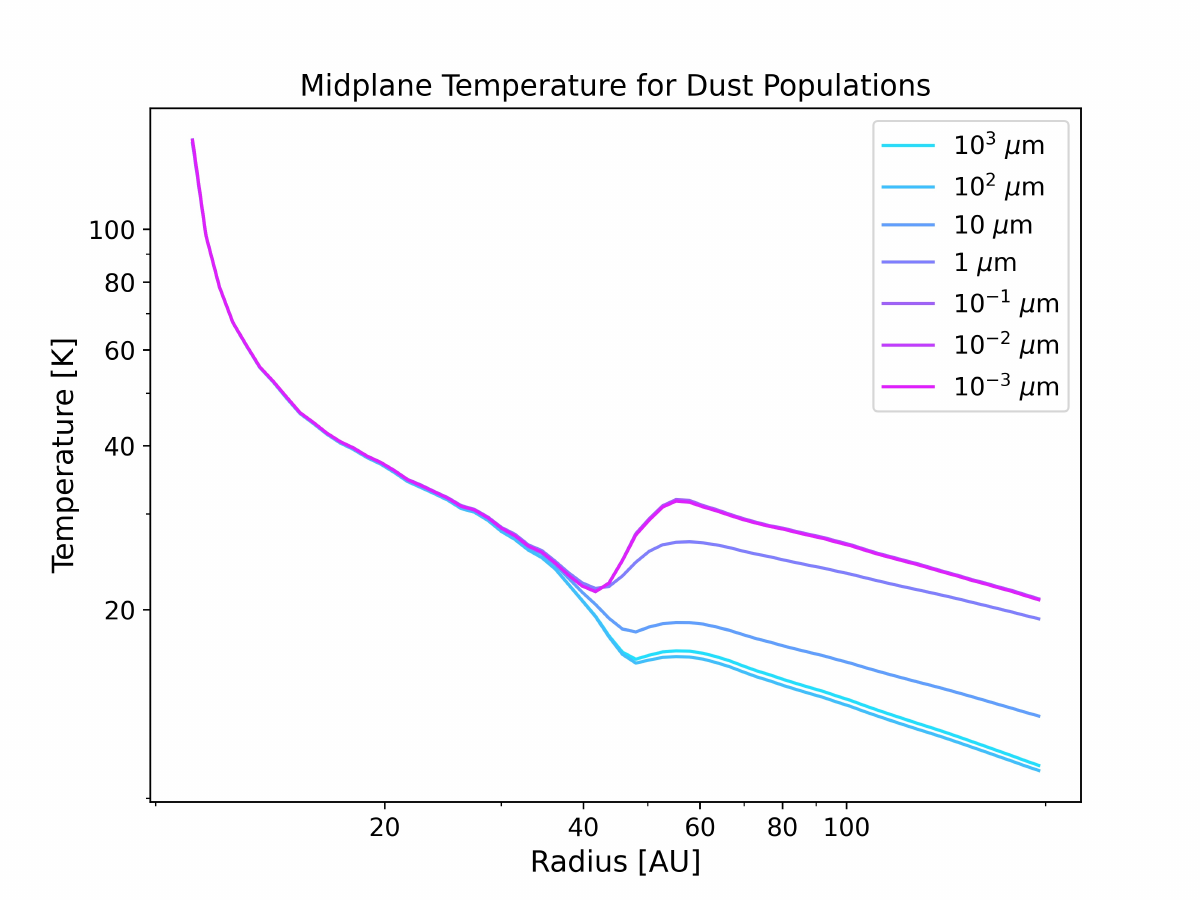}
         \caption{Initial midplane dust temperature for different species in the benchmark model (see Table \ref{table:1}).}
         \label{fig: T_mid_grain_sizes}
 \end{figure}

\begin{enumerate}
\item Starting from the initial disk setup, we first used \texttt{RADMC-3D} to compute the self-consistent initial dust temperature profile in the disk midplane. An example of the initial dust temperature profile from one of our models is shown in Fig.~\ref{fig: T_mid_grain_sizes}.
\item Assuming that the gas is thermally coupled to the smallest dust species, we derived the gas temperature profile and used it as input for the \texttt{DustPy} models of disk evolution.
\item After each time interval $t_{\rm iter}$, the evolved $\Sigma_{\rm d}$ and $\Sigma_{\rm g}$ profiles were passed back to \texttt{RADMC-3D} to update the dust temperature profile, after which the procedure returned to step 2.
\end{enumerate}

The iteration interval $t_{\rm iter}$ was set equal to the thermal relaxation timescale. Following the prescriptions in \citet{Zhang} and \citet{Bae_2021}, we estimated this timescale to be approximately 1000 yr at the location of the optically thin temperature bump in our models. Accordingly, we adopted $t_{\rm iter}$ = 1000 yr throughout the simulations.

\begin{figure*}
        \centering
        \includegraphics[width=\hsize]{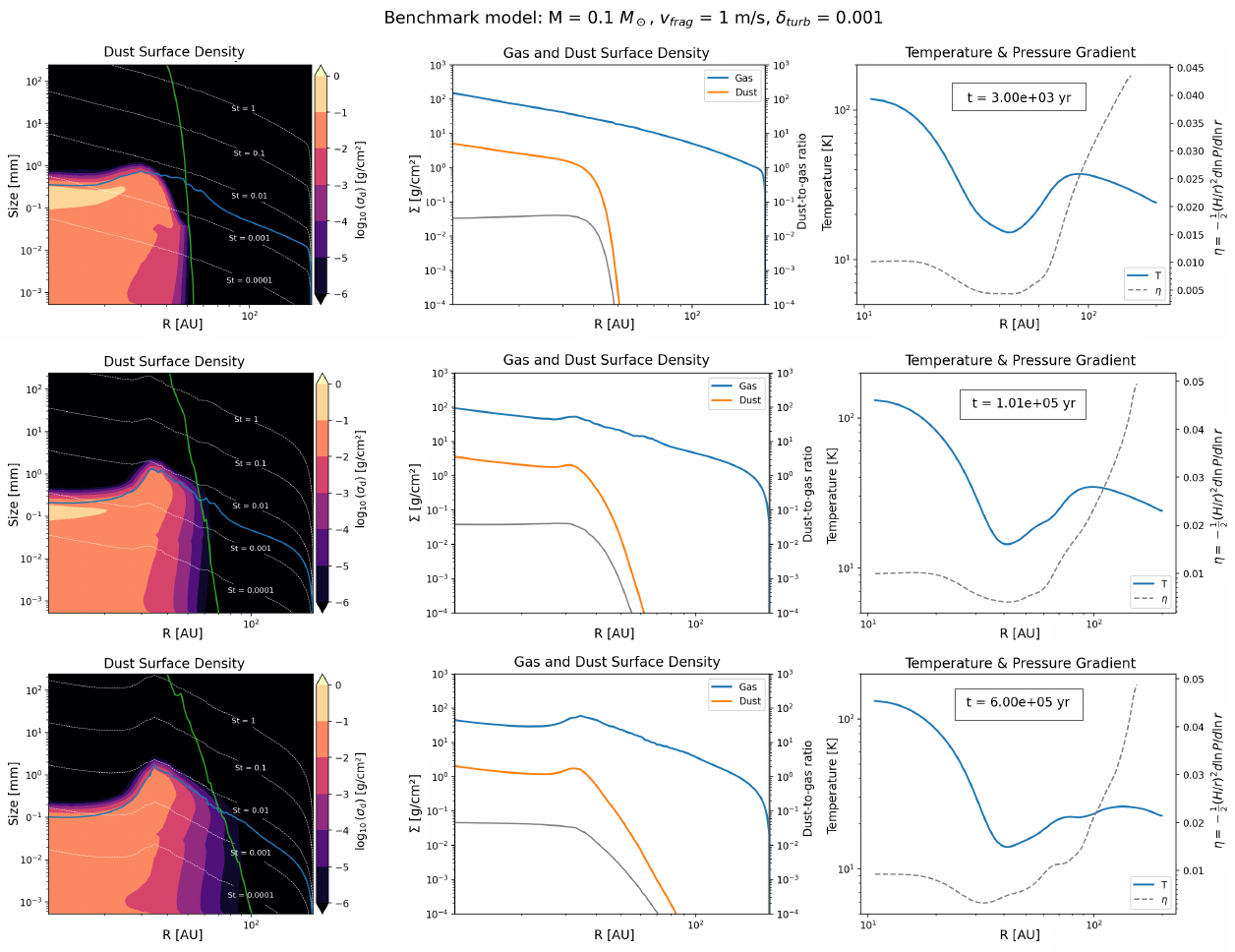}
        \caption{Snapshots for the benchmark simulation at 3000 yrs, 0.1 Myr and 0.6 Myr from top to bottom. {\bf Left panels} show the dust surface density as a function of both radius and grain size; the blue line is the fragmentation barrier, and the green line represents the drift barrier. {\bf Central panels} show the total dust surface density (orange), the gas surface density (blue), and the dust-to-gas ratio (gray) as a function of radius. {\bf Right panels} show the midplane temperature and the dimensionless pressure gradient $\eta$ as a function of radius.  
        }
        \label{fig: Benchmark_01_06}
\end{figure*}

\begin{table}[h!]
\caption{Summary of models conducted in this work.}                 
\label{table:1}    
\centering                        
\begin{tabular}{c c c c c}      
\hline\hline              
Name & Disk mass  & $v_{\text{frag}}$ & $\delta_{\rm t}$& $\delta_{\rm r}$  \\         
\hline                     
   Benchmark & 0.1 $M_{\odot}$ & 1 m/s &$10^{-3}$& $10^{-3}$ \\    
   High St & \textbf{0.01 $\bm{M_{\odot}}$} & \textbf{3 m/s}  & $10^{-3}$  & $10^{-3}$ \\
   Super High St & \textbf{0.001 $\bm{M_{\odot}}$} & \textbf{9 m/s}  & $10^{-3}$  & $10^{-3}$ \\
   High $\delta_{\rm t}$ & 0.1 $M_{\odot}$ & \textbf{3 m/s} & $\bm{10^{-2}}$  & $10^{-3}$ \\
   Low $\delta_{\rm t}$ & 0.1 $M_{\odot}$ & \textbf{0.3 m/s} &$\bm{10^{-4}}$& $10^{-3}$\\
   High $\delta_{\rm t},  \delta_{\rm r}$   &  $0.1 \ M_{\odot}$ & \textbf{3 m/s }&$\bm{10^{-2}}$   &  $\bm{10^{-2}}$\\
   Low $\delta_{\rm t},  \delta_{\rm r}$  & 0.1 $M_{\odot}$ & \textbf{0.3 m/s}&$\bm{10^{-4}}$   &  $\bm{10^{-4}}$ \\
   High St, high $\delta_{\rm t},  \delta_{\rm r}$ & $\bm{0.01 \ M_{\odot}}$ & \textbf{9 m/s} & $\bm{10^{-2}}$ & $\bm{10^{-2}}$  \\
   High St, low  $\delta_{\rm t},  \delta_{\rm r}$ & 
   $\bm{0.01 \ M_{\odot}}$ & 1 m/s & $\bm{10^{-4}}$ & $\bm{10^{-4}}$ \\
\hline     
\noalign{\smallskip}
Only dustpy & \multicolumn{4}{c}{ Temperature evolution turned off} \\
Extended & \multicolumn{4}{c}{ Added $30\%$ of extended small dust} \\
Different $n$ & \multicolumn{4}{c}{ Changing $n$ and $r_{\rm c}$ in the $\Sigma(r)$ profile}\\ 
\hline     

\noalign{\smallskip}
    
\end{tabular}
\end{table}

\begin{figure*}[h!]
        \centering
        \includegraphics[width=\hsize]{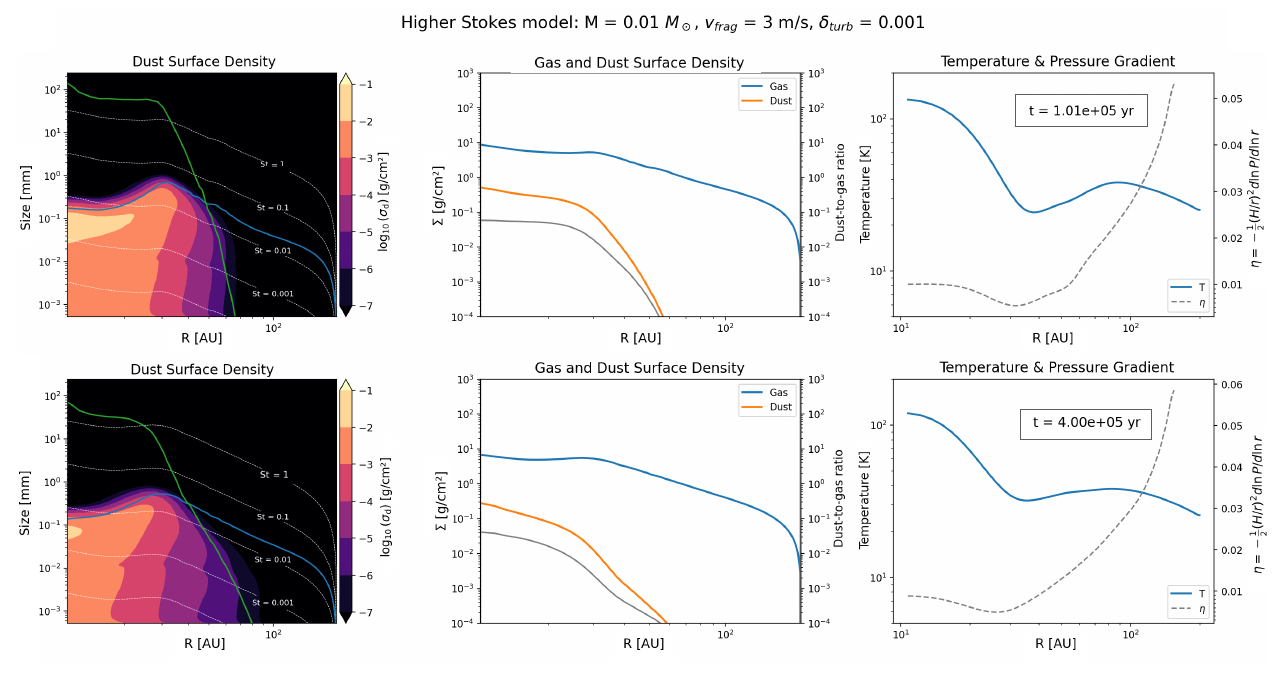}
        \caption{Snapshots for the higher Stokes number simulation at 0.1 Myr and 0.4 Myr from top to bottom. Notations are the same as Fig.~\ref{fig: Benchmark_01_06}.}
        \label{fig: Lighter_01_04}
\end{figure*}

\section{Results}
\label{sec: results}

\subsection{Benchmark model}
For the benchmark model, we adopted a disk mass $M = 0.1~M_{\odot}$, a fragmentation velocity $v_{\rm frag} = 1$~m~s$^{-1}$, and dust diffusion coefficients $\delta_{\rm t} = \delta_{\rm r} = 10^{-3}$. The simulation was evolved for 1~Myr. Figure~\ref{fig: Benchmark_01_06} shows snapshots at 0.003~Myr, 0.1~Myr, and 0.6~Myr.  
At early times ($t \lesssim 0.1$~Myr), the midplane temperature exhibits a local minimum followed by a bump beyond the outer edge of the dust disk. At the temperature minimum, the gas surface density is slightly higher because the kinematic viscosity in our $\alpha$-prescription depends on the sound speed, as in Eq.~\ref{eq: viscosity}. Since the sound speed decreases with temperature, the local temperature minimum leads to a lower viscosity. This in turn slows the radial mass transport down, causing gas to accumulate and producing a local enhancement in the surface density.

At the location of the temperature dip, the lower temperature reduces the thermal relative velocities of grains \citep{Ormel_2007},
\begin{equation}
\Delta v \sim c_{\rm s} \sqrt{\delta_{\rm t} \text{St}} ,
\label{eq: ormel}
\end{equation}
thereby elevating the fragmentation barrier. In other words, collisions are less energetic, allowing grains to grow to larger sizes before destructive collisions occur. The maximum Stokes number in the fragmentation-limited regime is
\begin{equation}
\text{St}_{\rm max} = \frac{v_{\rm frag}^2}{\alpha c_{\rm s}^2}, 
\end{equation}
and in the Epstein regime, the corresponding maximum particle size is
\begin{equation} 
a_{\rm max} \approx \frac{v_{\rm frag}^2 \Sigma_{\rm g}}{\rho_{\rm g}\alpha c_{\rm s}^2}.
\label{eq: maximum size}
\end{equation}
This explains why the largest grains preferentially form at the temperature minimum: the reduced sound speed allows particles to reach larger sizes, which in turn enhances cooling and helps maintain the temperature dip in a positive-feedback loop.
In Eq. \ref{eq: eta} we introduce the dimensionless pressure gradient parameter, $\eta$, which quantifies the deviation of the gas orbital velocity from Keplerian rotation due to radial pressure support. The radial drift velocity of dust is proportional to $\eta$ \citep{Birnstiel_review}, 
\begin{equation}
v_{\rm drift} \propto \frac{2r\eta \Omega_{\rm K}{\rm St} }{1+{\rm St}^2}
    \label{eq: v_drift}.
\end{equation}
Figure~\ref{fig: Benchmark_01_06} (right panels) shows the radial profile of $\eta$. Importantly, $\eta$ never reaches zero, indicating that dust grains are never completely halted at a fixed location. Instead, the dust accumulates in a traffic jam where the radial drift velocity varies with radius: inward drift slows at the gas density enhancement induced by the temperature dip, allowing a dust ring to form even though grains continue to drift inward. If the grains were too large (high Stokes numbers), their drift would be too fast to sustain the temperature dip and the associated gas pile-up, preventing ring formation. This motivates our exploration of models with larger Stokes numbers in Sect. \ref{sec:higher_St}.  
At later times ($t \sim 0.6$~Myr), the enhancement of the dust surface density becomes more pronounced. 

We ran the benchmark model, doubling the grid resolution and the number of photons, to perform a convergence test. We ran a model with 256 logarithmically spaced radial grid cells and $10^8$ photons. The resulting output was consistent with that of the 
benchmark model. This confirms that our model is numerically converged.

\subsection{Parameter study}

\subsubsection{Stokes number}
\label{sec:higher_St}
One of the most important parameters controlling the dust disk evolution is the Stokes number St. In the St$<1$ regime, in which our default model operates, a larger Stokes number means more efficient drift of dust along the pressure gradient.  
We first tested a model with a higher Stokes number. To control the experiments, we increased the fragment velocity from $v_{\text{frag}} = 1$~m/s to $v_{\text{frag}} = 3$~m/s and simultaneously decreased the disk mass to $M = 10^{-2}M_{\odot}$. This is lower by a factor of ten than the benchmark model. This design was employed to maintain the same maximum grain size under the fragmentation barrier (see Eq. \ref{eq: maximum size}). 
Because the gas surface density was ten times lower, the Stokes number of the particles was ten times higher than in the benchmark simulation, and therefore, the grains drift faster, as in Eq. \ref{eq: v_drift}.

Figure \ref{fig: Lighter_01_04} shows a snapshot of the higher-Stokes model simulation at 0.1 Myr and 0.4 Myr. 
In this model, the temperature dip and bump do not represent a stable configuration, and by 0.4 Myr, the minimum has clearly become shallower and the bump has decreased in magnitude. The fragmentation barrier has a maximum at the location of the temperature minimum, and particles can grow to larger sizes, but the larger particles rapidly drift inward.
To sustain enhanced grain growth and maintain the temperature dip, the larger grains (the main cooling source) must remain in this region long enough. In the higher-Stokes model, radial drift prevents the grains from lingering at the dip location. As a result, no ring forms, and the disk profile is smooth after 1 Myr.
We ran a model with an even higher Stokes number and decreased the mass to 0.001 $M_{\odot}$ and increased $v_{\text{frag}} = 9 \ \hbox{m/s}$. In this model, the grains drifted even faster, and the temperature profile was monotone by $0.1 \ \hbox{Myr}$.

\subsubsection{Mutual collision of particles}

In \texttt{Dustpy}, the turbulent relative velocity between dust grains is regulated by the turbulent diffusivity parameter $\delta_{\rm t}$, according to Eq. \ref{eq: ormel}, which affects the collisional evolution of the dust. To directly assess the role of this parameter, we ran two models: a high-turbulence case with $\delta_{\rm t} = 10^{-2}$, and a low-turbulence case with $\delta_{\rm t} = 10^{-4}$. We point out that this is only the dust turbulent coefficient; the gas is not controlled by it. In both simulations, the fragmentation velocity was adjusted consistently following Eq.~\ref{eq: maximum size} in order to preserve the maximum grain size. We therefore compared models with different collision rates.
\begin{figure}
    \centering
    \includegraphics[width=\hsize]{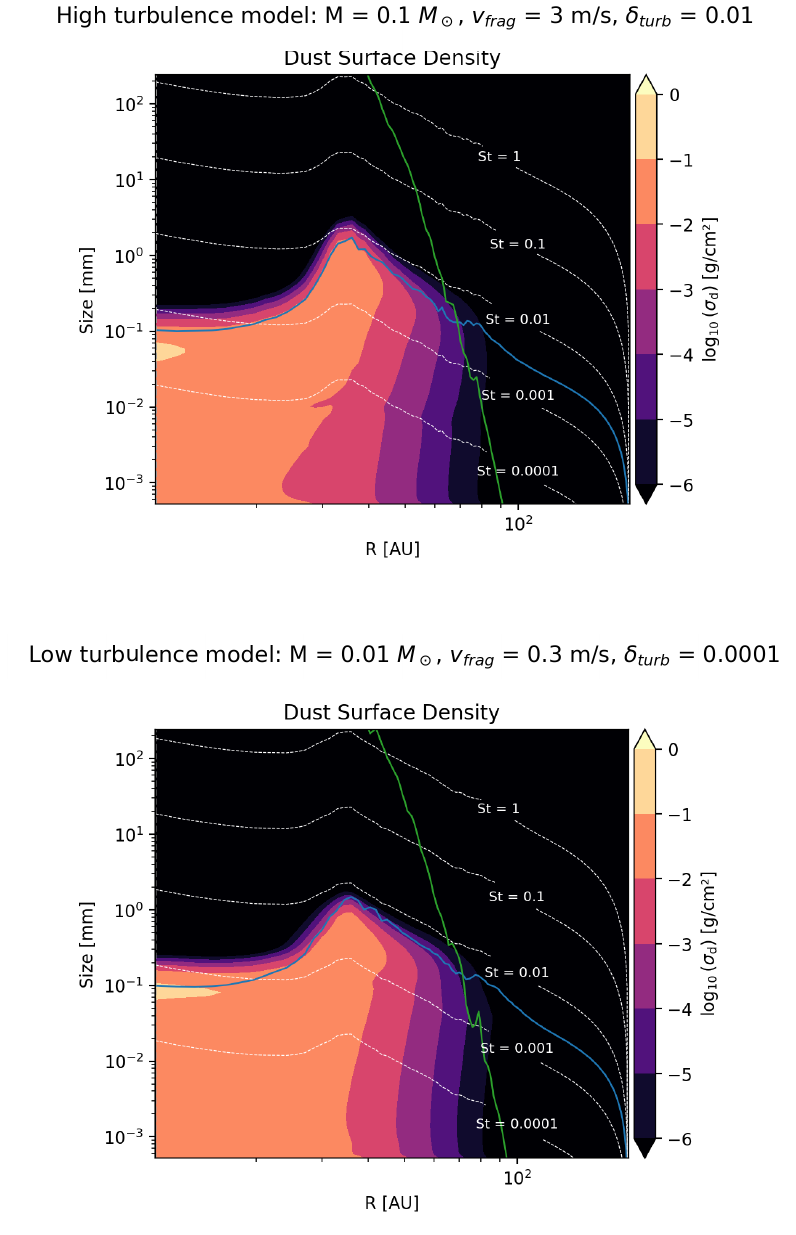}
    \caption{\textit{Top}: Snapshot of the high $\delta_{\rm t}$ model after 0.5 Myr. \textit{Bottom}: Snapshot of the low $\delta_{\rm t}$ model after 0.5 Myr. The notation is the same as in the left panel of Fig. \ref{fig: Benchmark_01_06}.}
    \label{fig:delta_t}
\end{figure}
Figure \ref{fig:delta_t} shows two snapshots taken at 0.5 Myr for the high- and low-turbulence simulations. In both cases, a density enhancement forms at the disk edge, indicating that $\delta_{\rm t}$ does not alter the overall structure strongly. However, in the high-turbulence case, the ring has a thinner peak, while in the low-turbulence case, it is smoother. This behavior can be understood by noting that higher values of $\delta_{\rm t}$ correspond to more frequent and energetic collisions, which promote grain growth.

\subsubsection{Diffusion and turbulence}
\begin{figure}
    \centering
    \includegraphics[width=0.9\linewidth]{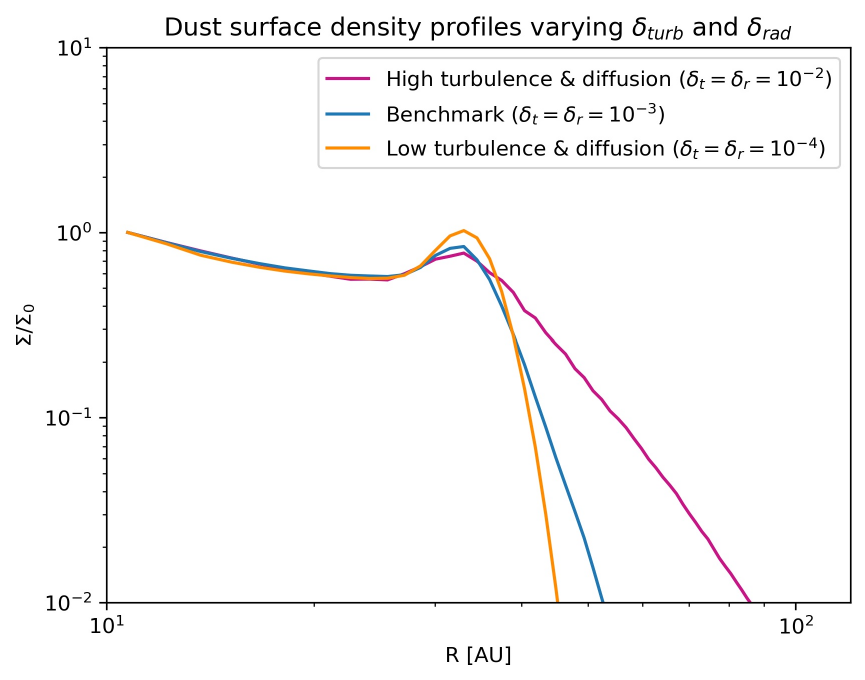}
    \caption{Dust surface density at 0.3 Myr varying $\delta_{\rm r}$ and $\delta_{\rm t}$ in the models.}
    \label{fig:delta_t_delta_r}
\end{figure}

The radial dust diffusion is regulated by the diffusion coefficient as in Eq. \ref{eq: dust_diffusion}, where $\delta_{r}$ is the radial dust diffusivity parameter.
To explore the combined effects of turbulence and diffusion, we performed two additional simulations in which $\delta_{\rm t}$ and $\delta_{\rm r}$ were varied simultaneously. The fragmentation velocity $v_{\mathrm{frag}}$ was adjusted accordingly to keep the maximum grain size fixed. The high diffusion and turbulence model adopted $\delta_{\rm t} = \delta_{\rm r} = 10^{-2}$, while in the low diffusion and turbulence model, the two parameters were set to $10^{-4}$.
Figure \ref{fig:delta_t_delta_r} presents a snapshot of the surface dust densities taken at 0.3 Myr for both models, with the benchmark model ($\delta_{\rm t} = \delta_{\rm r} = 10^{-3}$). The concentration of the surface density peak increases as the radial diffusion decreases. The results indicate that $\delta_{\rm r}$ is the dominant parameter in shaping the appearance of the ring with respect to $\delta_{\rm t}$. The lower $\delta_{\rm r}$, the thinner and more concentrated the density enhancement; when $\delta_{\rm r}$ is increased, the peak becomes broader and more diffuse. This is the result of strong diffusion smoothing the disk edge and making it less sharp. As a consequence, the temperature transition at the edge becomes more gradual, reducing the efficiency of dust growth and leading to a less highly concentrated dust enhancement.

\subsubsection{Changing both the Stokes number and the diffusion coefficients}
\begin{figure}
    \centering
    \includegraphics[width=\hsize]{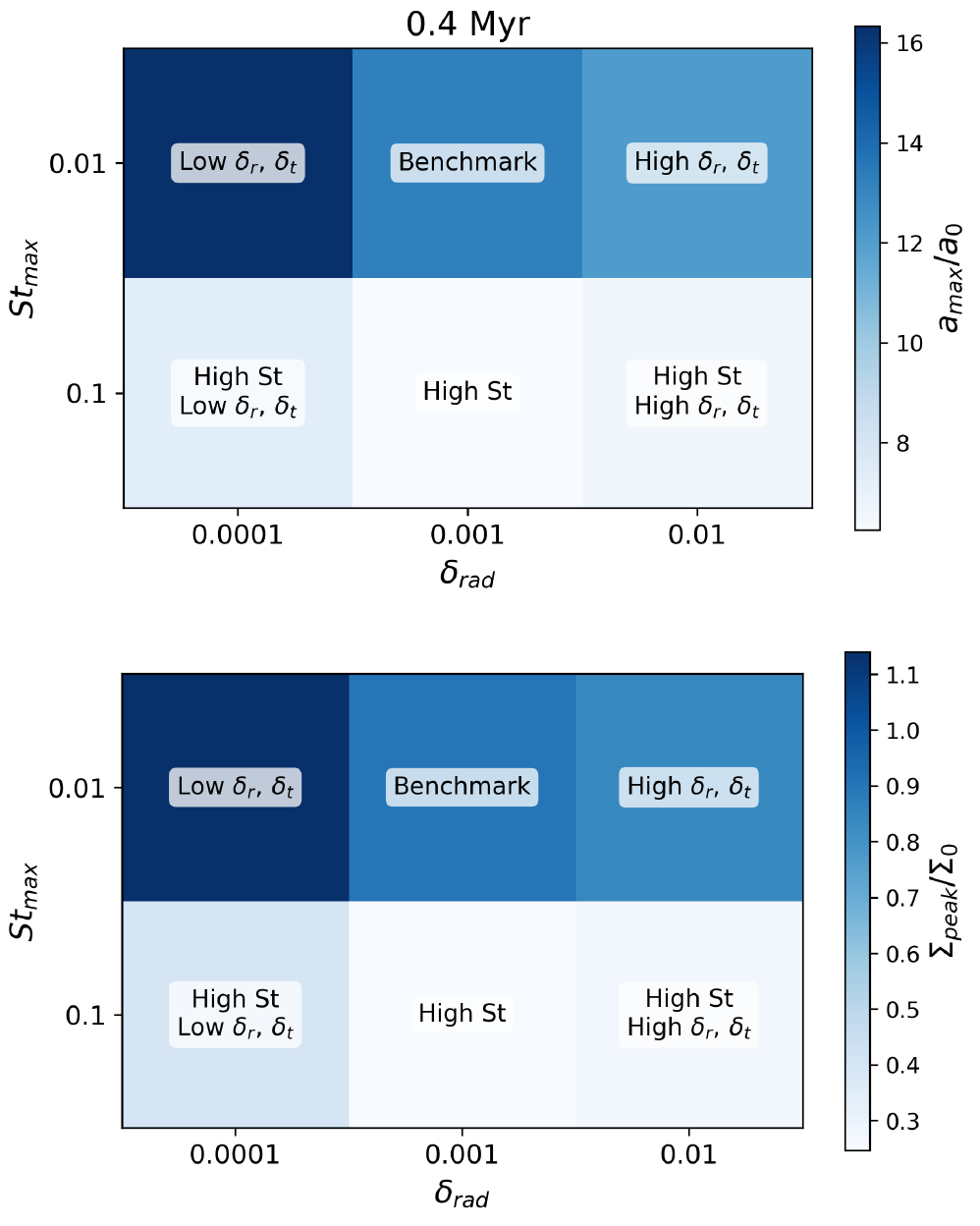}
    \caption{Normalized maximum grain size $a_{\text{max}}/a_0$ and normalized peak dust surface density $\Sigma_{\text{peak}}/\Sigma_0$ for different combinations of Stokes number and dust diffusivity parameters. $a_0$ and $\Sigma_0$ are respectively the grain size and the dust surface density at the inner radius. The maximum grain size and the peak surface density correspond to the temperature dip location.}
    \label{fig:table}
\end{figure}

We ran two additional models in which the Stokes number and the diffusion coefficients were varied simultaneously because these two quantities affect the radial motion of particles. In both models, we set the disk mass to $M=0.01 \  M_{\odot}$, resulting in particles with a Stokes number that was larger by one order of magnitude than in the benchmark model, as discussed in Sect.~\ref{sec:higher_St}. We varied the diffusion parameters as follows:
\begin{itemize}
    \item High Stokes number, high-diffusion model with $\delta_{\rm r} = \delta_{\rm t} = 10^{-2}$ and $v_{\text{frag}}=9 \hbox{ m/s}$;
    \item High Stokes number, low-diffusion model with $\delta_{\rm r} = \delta_{\rm t} = 10^{-4}$ and $v_{\text{frag}}=1 \hbox{ m/s}$.
\end{itemize}
In neither case, a ring formed because the higher Stokes number leads to rapid inward drift of the largest grains. However, Fig.~\ref{fig:table} provides a useful comparison by showing the maximum grain size $a_{\text{max}}$ measured at the location of the temperature dip and normalized to the grain size at the inner disk edge $a_0$, together with the normalized dust surface density peak $\Sigma_{\text{peak}}/\Sigma_0$ for all models exploring variations in Stokes number and dust diffusion.
The contrast in the top row, corresponding to the benchmark Stokes number (St =0.01), is significantly higher than in the bottom row. In this regime, $a_{\text{max}}/a_0$ and $\Sigma_{\text{peak}}/\Sigma_0$ decrease systematically with increasing radial diffusivity $\delta_{\rm r}$. The same qualitative trend with the radial diffusivity is observed in models with a higherStokes number, although in this case, the overall contrast is reduced and no distinct ring structure develops.

\subsubsection{Without radiative transfer}
To confirm that the emergence of the ring is due to the temperature structure, we ran a control model in which the disk was evolved only with \texttt{DustPy}, starting from the initial density profile given in Eq.~\ref{eq: density_profile}, but neglecting radiative transfer. In this case, the disk evolution remained smooth, confirming that the ring formation seen in the full model arises from the non-monotonic temperature structure.
\subsubsection{Adding extended dust}
We investigated whether, as first proposed by \citet{CrCha}, the temperature bump at the disk edge might generate a corresponding pressure bump, since
\begin{equation}
    P_{\text{mid}}(r) = \frac{\rho_{\text{mid}}(r)\ T_{\text{mid}}(r)\ k_B}{\mu \ m_H},
\end{equation}
and whether such a pressure bump could act as a dust trap. In our basic setup, the temperature bump sat just outside the disk edge and was therefore ineffective. To trap a significant amount of dust, strong outward dust diffusion is required. However, if this were to be the case, this process would simultaneously smooth the disk edge, thereby reducing the prominence of the bump. 

\begin{figure}
    \centering
    \includegraphics[width=\hsize]{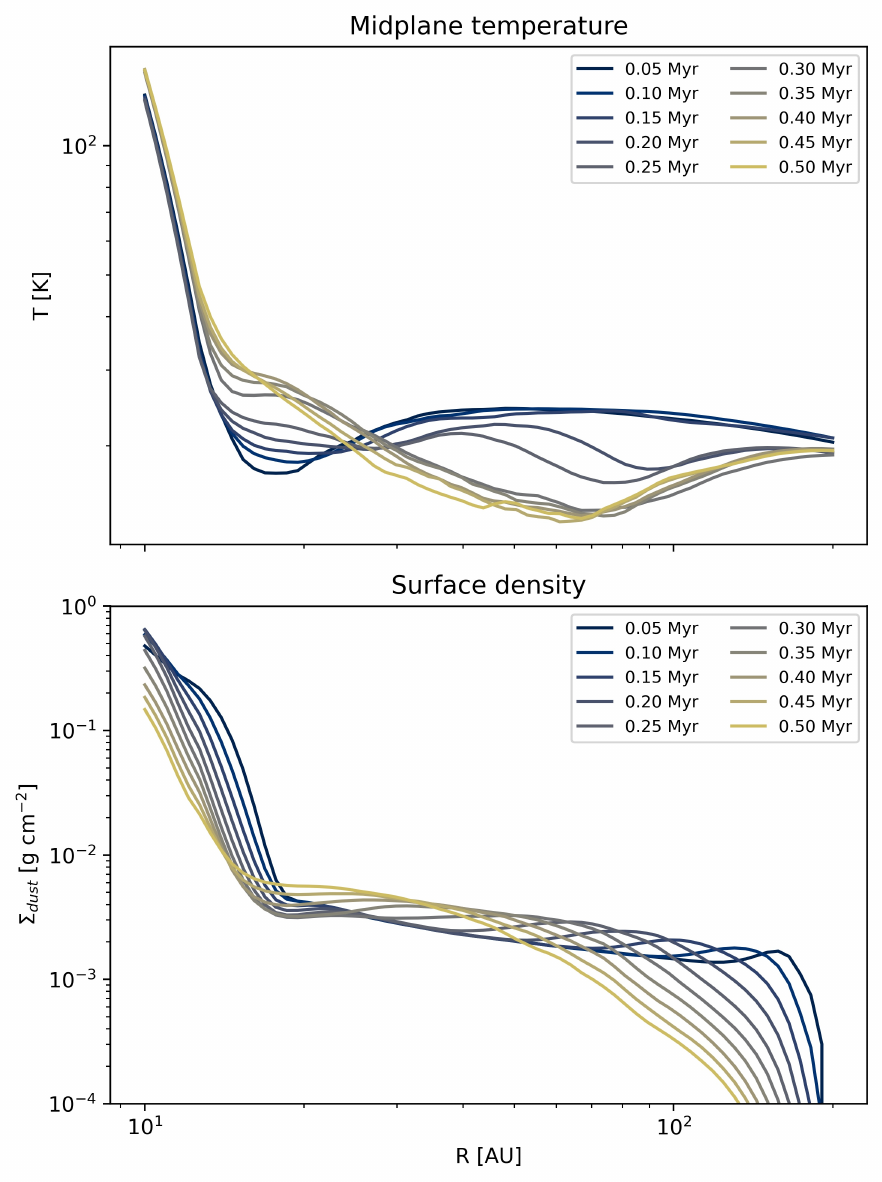}
    \caption{Time evolution from 0 to 0.5 Myr of the midplane temperature (top panel) and dust surface density (bottom panel) of the model with added 30\% of mass in small extended grains.}
    \label{fig:Adding_extended}
\end{figure}

To test whether the bump might trap dust under more favorable conditions, we performed an alternative setup in which we added an extended dust component: we included an additional $30\%$ of the dust mass in small $0.1\ \mu$m grains in addition to the standard dust distribution, spatially distributed as the gas, like following a late in-fall.  
In Fig. \ref{fig:Adding_extended} we show that by 0.5 Myr, the dust growth front in the outer disk reaches the bump location, where grains rapidly grow to larger sizes. These larger grains reduce the local temperature and erase the bump. Our result indicates that a temperature bump cannot create a stable pressure bump, since when grains start to grow within the bump, the cooling effect of larger grains suppresses the temperature enhancement, preventing the formation of a long-lived dust trap.

\subsubsection{The sharpness of the edge and the disk size}
The profile in Eq. \ref{eq: density_profile} is peculiar because of the exponent $n=10$ in the exponential taper part. \citet{Birnstiel_outer_edge} showed that in the early phases of disk evolution, radial drift (and to a lesser extent, gas drag) naturally produces a sharp outer dust edge. A similarly steep edge can also result from tidal truncation by a companion, such as a binary star or a massive planet \citep[e.g.,][]{Artymowicz_1994}.
However, it seems arbitrary and unlikely that a disk should have exactly this profile in the early stages of its lifetime. In Appendix \ref{sec: appendix} we show that there is a degeneracy between $n$ and $r_{\rm c}$, and we can obtain a temperature bump at the same location with different $n$ by varying $r_{\rm c}$: the lower the exponent, the smaller the cut-off radius. \citet{Curone_exoALMA} fit the exponential decline in the intensity of the emission of their sample between $R_{90}$  and $R_{\text{out}}$ such that the values of the intensity were below five times the rms noise $\sigma_{\text{rms}}$. 
They generally imposed $n=1$ and found $r_{\rm c}$ generally on the order of one tenth of $R_{\text{out}}$: around $10$ AU for disks detected until 100 AU.
We ran our simulation and varied $n$ and $r_{\rm c}$ according to Eq. \ref{eq:change_rc} in order to obtain the bump at approximately the same location as the benchmark model. We obtained the ring for each exponent $n$.

\subsection{Intensity profile of the ring}
\begin{figure}[h!]
        \centering
        \includegraphics[width=0.8\hsize]{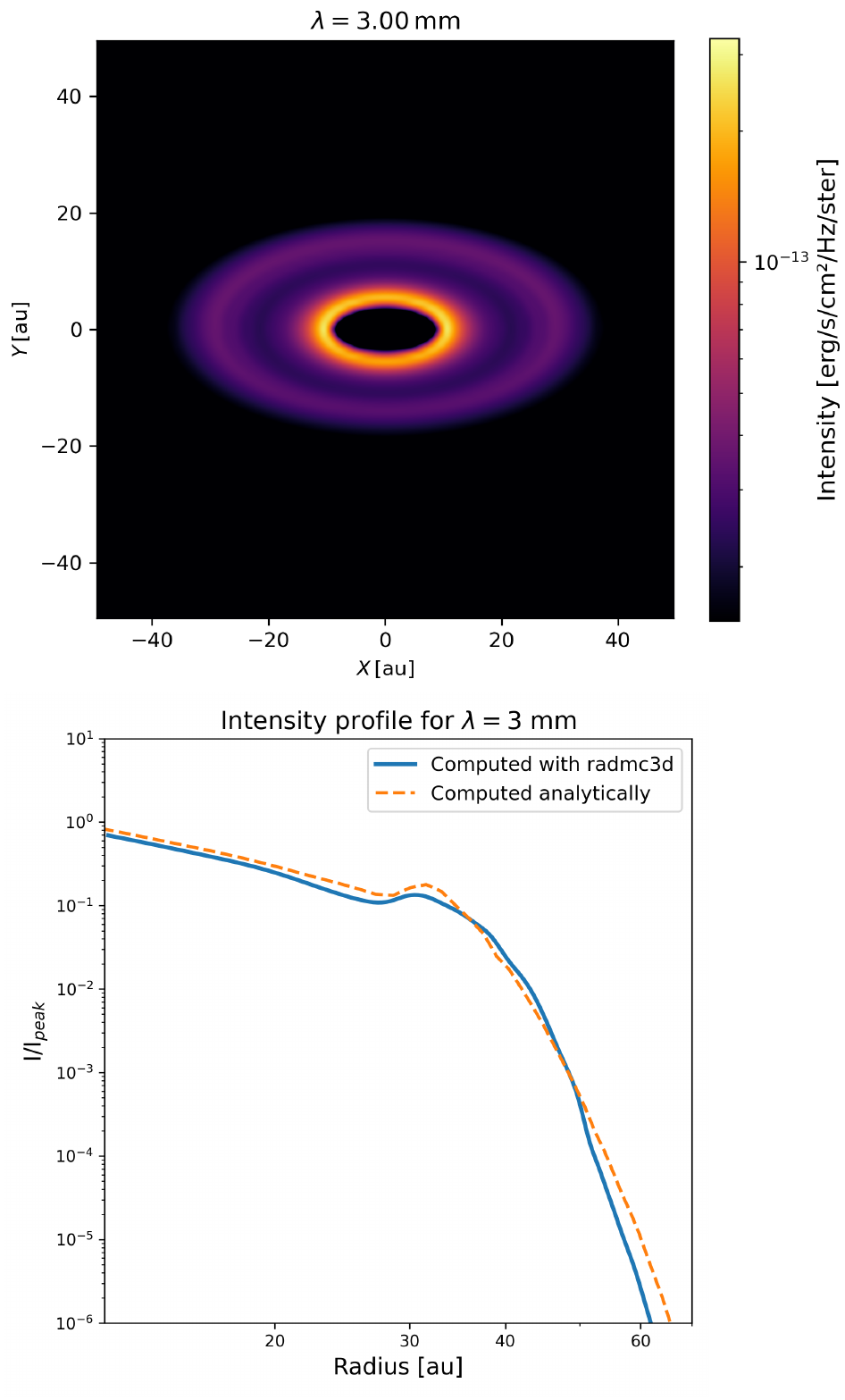}
        \caption{Top panel: image computed at 3 mm with \texttt{RADMC-3D}. Bottom panel: normalized radial intensity profile computed analytically following Equation \ref{eq: intensity}, or numerically from the \texttt{RADMC-3D} image.}
        \label{fig: intensity_profiles}
\end{figure}
In order to assess whether this mechanism would produce an 
observable ring, we generated a synthetic intensity map of the i
emission. An analytic estimate of the intensity profile can be 
obtained from the \texttt{DustPy} output as follows. Given the 
dust surface density as a function of radius and grain size 
$\Sigma_{\text{dust}}(r,a)$ and the opacity of each grain size 
$\kappa_{\nu}(a)$, the total opacity as a function of radius is 
obtained by weighting over the grain size distribution,
\begin{equation}
\kappa_{\nu}(r) = \frac{\sum_a \kappa_{\nu}(a)\,
\Sigma_{\text{dust}}(r,a)}{\sum_a \Sigma_{\text{dust}}(r,a)}.
\end{equation}
In the optically thin approximation, the intensity is then
\begin{equation}
I_{\nu}(r) = B_{\nu}(T(r))\,\kappa_{\nu}(r)\,\Sigma_{\text{dust}}(r)
\label{eq: intensity}
,\end{equation}
where $\Sigma_{\text{dust}}(r) = \sum_a \Sigma_{\text{dust}}(r,a)$ 
is the total dust surface density.

To obtain a more accurate estimate, we also computed the intensity 
profile with \texttt{RADMC-3D}. From the final \texttt{DustPy} 
output, we obtained $\Sigma_{\text{dust}}(r,a)$, and from the 
preceding iteration, the temperature profile $T(r)$. The gas scale 
height was then computed as $H = \sqrt{k_B T(r) / (m_H \mu)}$, and 
the dust scale height for each grain size $a$ was obtained by 
balancing vertical diffusion and settling \citep{Dubrulle, 
Birnstiel2010} as in Eq. \ref{eq: scale_height}.
This procedure yielded the 3D dust density 
distribution required as input for \texttt{RADMC-3D}. We computed 
a synthetic image at $\lambda = 3$~mm using $10^8$ photons and 
derived the radial intensity profile by averaging over the azimuthal 
coordinate $\phi$. Figure~\ref{fig: intensity_profiles} shows 
the analytic and numerical intensity profiles, both of which 
exhibit an enhancement before declining. This is consistent with an edge 
ring.

\subsection{Potential hydrodynamical instabilities}
We verified whether our disk model was susceptible to two instabilities: the Rossby wave instability, driven by the radial disk structure, and the streaming instability, driven by the local dust-to-gas ratio and grain size.
\subsubsection{Rossby wave instability}
The Rossby wave instability \citep{Lovelace_1999} is triggered by steep variations in the density or entropy profile of a disk. The instability criterion is to have an extremum in the function $\mathcal{L}(r)$, a quantity related to the vorticity of the equilibrium flow. Defining $S\equiv P/\Sigma^{\Gamma}$ as the entropy of the disk matter, $\Omega$ as the orbital frequency, and $\kappa$ as the epicyclic frequency, we write $\mathcal{L}$ as
\begin{equation}
    \mathcal{L}(r)\approx \frac{\Sigma\,\Omega\, S^{2/\Gamma}}{\kappa^2}
    \propto \frac{\Omega}{\kappa^2}\,\Sigma^{(2-\Gamma)/\Gamma}\,T^{2/\Gamma}.
\end{equation}
The analytical instability criterion requires $d\mathcal{L}/dr=0$; when this condition is satisfied,  linear perturbations are expected to grow \citep{Meheut_2010, Chang_2023}. Because the exponent of $T$ is generically larger than that of $\Sigma$ for typical values of $\Gamma$, a variation in temperature has a stronger effect on $\mathcal{L}$ than an equivalent fractional variation in surface density \citep{Lovelace_1999}.
In our disk, two radii satisfy the extremum condition of the Lovelace instability criterion, corresponding to the locations of the temperature minimum (giving a minimum in $\mathcal{L}$) and the temperature maximum (giving a maximum in $\mathcal{L}$).
For the maximum, we refer to \citet{Chang_2023}, who computed the stability of a disk against RWI for a Gaussian temperature bump as a function of its amplitude and width. Our temperature maximum is not exactly Gaussian, but fitting it with a Gaussian profile and comparing to their results indicates that it is RWI unstable. 
However, the temperature maximum lies outside the dust disk, and modeling the effect of RWI at this location on the dust distribution is beyond the scope of this paper.
Regarding the minimum, \citet{Lovelace_1999} showed analytically that perturbations 
at a temperature minimum propagate away from the extremum and are not trapped, 
in contrast to the maximum case. This suggests that while the necessary condition 
for RWI is formally satisfied, sustained exponential growth of a coherent unstable 
mode is unlikely to develop at this location without a proper trapping mechanism. 
A full numerical eigenvalue analysis is required to confirm this.
\begin{figure}
    \centering
    \includegraphics[width=0.9\hsize]{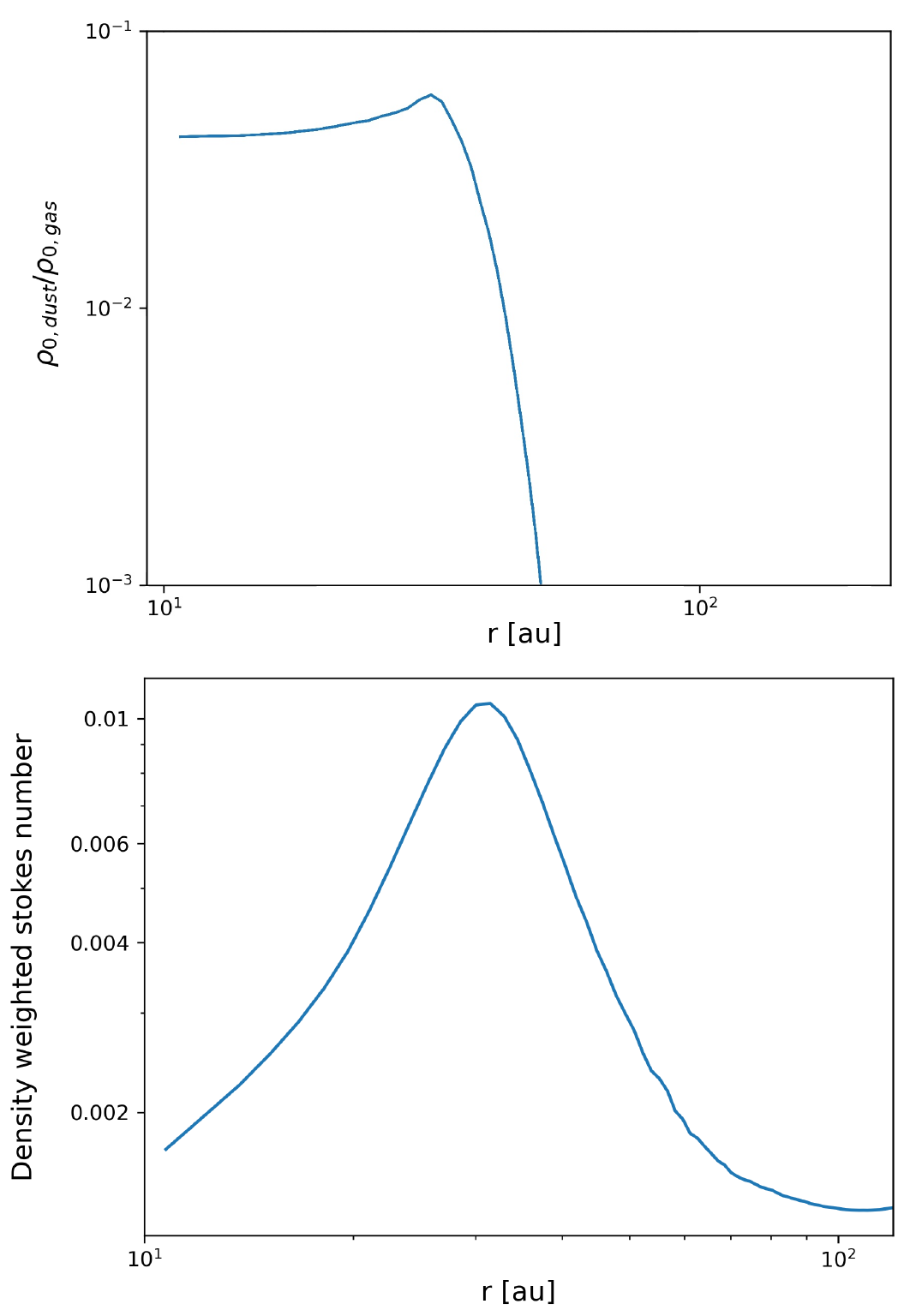}
    \caption{\textit{Top}: Midplane dust-to-gas ratio for the low-turbulence and diffusion model at 0.6 Myr. \textit{Bottom}: Density-weighted Stokes number for the low-turbulence and diffusion model at 0.6 Myr}
    \label{fig:SI}
\end{figure}
\subsubsection{Streaming instability}
We tested whether the dust ring was unstable to the streaming instability (SI).
The SI can concentrate solids into dense clumps, and it is widely regarded as a leading mechanism for the formation of planetesimals \citep{Youdin_Goodman_2005}. To assess the dust-to-gas ratio conditions at the ring more quantitatively, we computed $\varepsilon_{\rm mid}(R)$ for the low-turbulence low-diffusion model, which produces the most contrasted ring among our models (Fig.~\ref{fig:delta_t_delta_r}).
The midplane dust-to-gas ratio is higher at the ring location, reaching a peak value of $\varepsilon_{\rm mid}\approx6\times10^{-2}$ (top panel of Fig.~\ref{fig:SI}). We also computed the density-weighted Stokes number as a function of radius (bottom panel of Fig.~\ref{fig:SI}),
\begin{equation}
\langle \mathrm{St} \rangle(r) =
\frac{\displaystyle \int_{a_{\min}}^{a_{\max}}
\mathrm{St}(a,r)\,\rho_{\mathrm{d}}(a,r)\,\mathrm{d}a}
{\displaystyle \int_{a_{\min}}^{a_{\max}}
\rho_{\mathrm{d}}(a,r)\,\mathrm{d}a}
,
\end{equation}
where the sum runs over the dust size bins $i$, and $\rho_{{\rm d},i}$ is the midplane dust density of the $i$th size bin.
This quantity likewise peaks at the ring location, reaching $\mathrm{St}\approx1.1\times10^{-2}$ and decreases inward and outward from the ring. 
Comparing the peak $\varepsilon_{\rm mid}$ to the runaway clumping criterion, $\varepsilon_{\rm mid}\gtrsim1$ \citep{Johansen_Youdin_2007}, we found that in the most favorable of our models, the midplane dust-to-gas ratio remains almost an order of magnitude below this threshold even at the ring. This threshold was found to be lower for grains with $\mathrm{St}\gtrsim0.01$--$0.02$, for which \citet{Li_Youdin_2021} reported a critical ratio of $\varepsilon_{\rm crit}\approx0.35$--$1$ that sharply increased below this range. The density-weighted Stokes number we computed, $\mathrm{St}\approx0.01$, sits right at the lower edge of this range, where the critical threshold is already close to its highest values. We note that this comparison, based on the density-weighted St, is only an optimistic proxy for a genuinely monodisperse population, since the \citet{Li_Youdin_2021} thresholds were derived for single-species dust, whereas our disk retained a full grain size distribution at the ring. A true polydisperse population is expected to be more stable than this monodisperse estimate suggests \citep{Krapp_2019, Paardekooper_2020}. Even under this optimistic assumption, however, $\varepsilon_{\rm mid}$ remains well below $\varepsilon_{\rm crit}$. We therefore conclude that the streaming instability is unlikely to be triggered at the ring under the conditions of our current models.

However, we note that our models did not account for any dust feedback on gas, which might promote dust settling \citep{xu.Z.2022.01} and further slow down the radial drift at the ring location \citep{nakagawa.Y.1986.09}. Inclusion of dust feedback might help trigger a streaming instability and is a promising follow-up to our work. 

\section{Discussion and conclusions}
\label{sec: conclusions}
We investigated a possible mechanism for the 
formation of edge rings in protoplanetary disks through thermodynamic feedback arising at the outer dust edge. We modeled this process by 
coupling dust transport and grain-growth simulations performed with 
\texttt{DustPy} to radiative transfer calculations carried out 
with \texttt{RADMC-3D}.

A sharp dust surface density gradient at the outer disk 
edge produced a non-monotonic temperature structure. Beyond the dust edge, the optical depth to stellar radiation decreases, allowing 
stellar photons to penetrate more deeply and heat the midplane. 
At the same location, large dust grains, which contribute more efficiently to radiative cooling, are absent, reducing the local cooling efficiency. 
The combination of enhanced heating and reduced cooling produces a temperature bump just beyond the outer dust edge, which in turn modifies 
dust growth and radial transport.

We studied the evolution of a disk with a steep outer edge, 
described by a power-law surface density profile with an exponential taper 
(Eq.~\ref{eq: density_profile}), using our iterative 
radiative transfer and dust-evolution framework 
(Fig.~\ref{fig: code_workflow}). Our main findings are listed below.

\begin{itemize}
    \item A disk with a steep outer edge naturally develops a non-monotonic 
    temperature profile, characterized by a dip followed by a bump 
    near the location where the optical depth to stellar irradiation 
    reaches $\tau_\star = 2/3$. This behavior is robust across a 
    range of dust compositions and density profiles.
    
    \item When dust evolution is included, the temperature bump 
    does not produce a stable pressure trap, as suggested by \citet{kim.S.2020.01}. The reason is that grain growth within 
    the bump enhances radiative cooling and suppresses the temperature 
    enhancement itself.
    
    \item At the temperature minimum (the dip), the lower sound speed raises the fragmentation barrier, allowing grains to grow to larger sizes. 
    These larger grains further enhance radiative cooling, reinforcing the temperature minimum through a positive thermodynamic feedback loop.
    
    \item Through the adopted $\alpha$-viscosity prescription, the temperature dependence of the viscosity leads to a 
    local gas pile-up at the temperature minimum.
    
    \item The grain growth driven by the temperature dip produces 
    an observable ring-like continuum emission enhancement (Fig.~\ref{fig: 
    intensity_profiles}), which persists across a range of 
    turbulence parameters and surface density profile slopes.
    
    \item Higher Stokes numbers increase the radial drift, causing dust to leave the disk edge more rapidly and suppressing the formation of the dust enhancement.
    
    \item The parameter $\delta_{\rm t}$, which regulates the turbulent relative velocities and collision rates between particles, affects the overall structure only little, although higher values of $\delta_{\rm t}$ produce narrower and more strongly peaked ring dust enhancements.
    
    \item The radial diffusion coefficient $\delta_{\rm r}$ strongly affects the dust concentration. Higher values of 
    $\delta_{\rm r}$ smooth the outer disk edge and produce broader, less concentrated dust enhancements.

\end{itemize}

Although the dust enhancements produced in our models do not correspond to long-lived pressure traps, they may nevertheless have important implications for planet formation. By slowing radial drift and locally increasing the dust-to-gas ratio, these edge-ring structures might provide an initial site for dust accumulation in the outer disk. Such enhancements can facilitate the formation of planetesimals through collective processes such as the streaming instability \citep[e.g.,][]{xu.Z.2022.01,lim.J.2025.03}, after which subsequent growth through pebble accretion can proceed efficiently \citep{jiang.H.2023.01}. In this picture, thermally induced edge rings can represent a very early stage of planet formation and can help us to explain the origin of wide-orbit protoplanets in young disks \citep[e.g.,][]{keppler.M.2018.09,vancapelleveen.RF.2025.08}. This potentially helps to alleviate the chicken-and-egg problem of forming the first generation of planets.

These results suggest that temperature-driven effects can naturally 
explain the edge rings observed in protoplanetary 
disks. Our 
study further underscores the importance of self-consistently 
coupling thermodynamics and dust evolution when modeling disk 
substructures \citep[see, e.g.,][]{Zhang, Zheao_2024, Dhruv_2024, 
Chen_2025}.

\begin{acknowledgements}
    We thank the referee for the comments that improved the quality of the manuscript. We thank Carlo Nipoti for the useful discussion about hydrodynamical instabilities. H.J. acknowledges support by the German Aerospace Center (DLR) and the Federal Ministry for Economic Affairs and Energy (BMWi) through program 50OR2513 ‘RAP’.  J.B. was supported by the Deutsche Forschungsgemeinschaft (DFG) under Grant No. 544937803. 
\end{acknowledgements}

\bibliographystyle{aa}
\bibliography{biblio}

\begin{appendix}
\section{Temperature bump model}
\label{sec: appendix}
\begin{figure}
    \centering
    \includegraphics[width=\hsize]{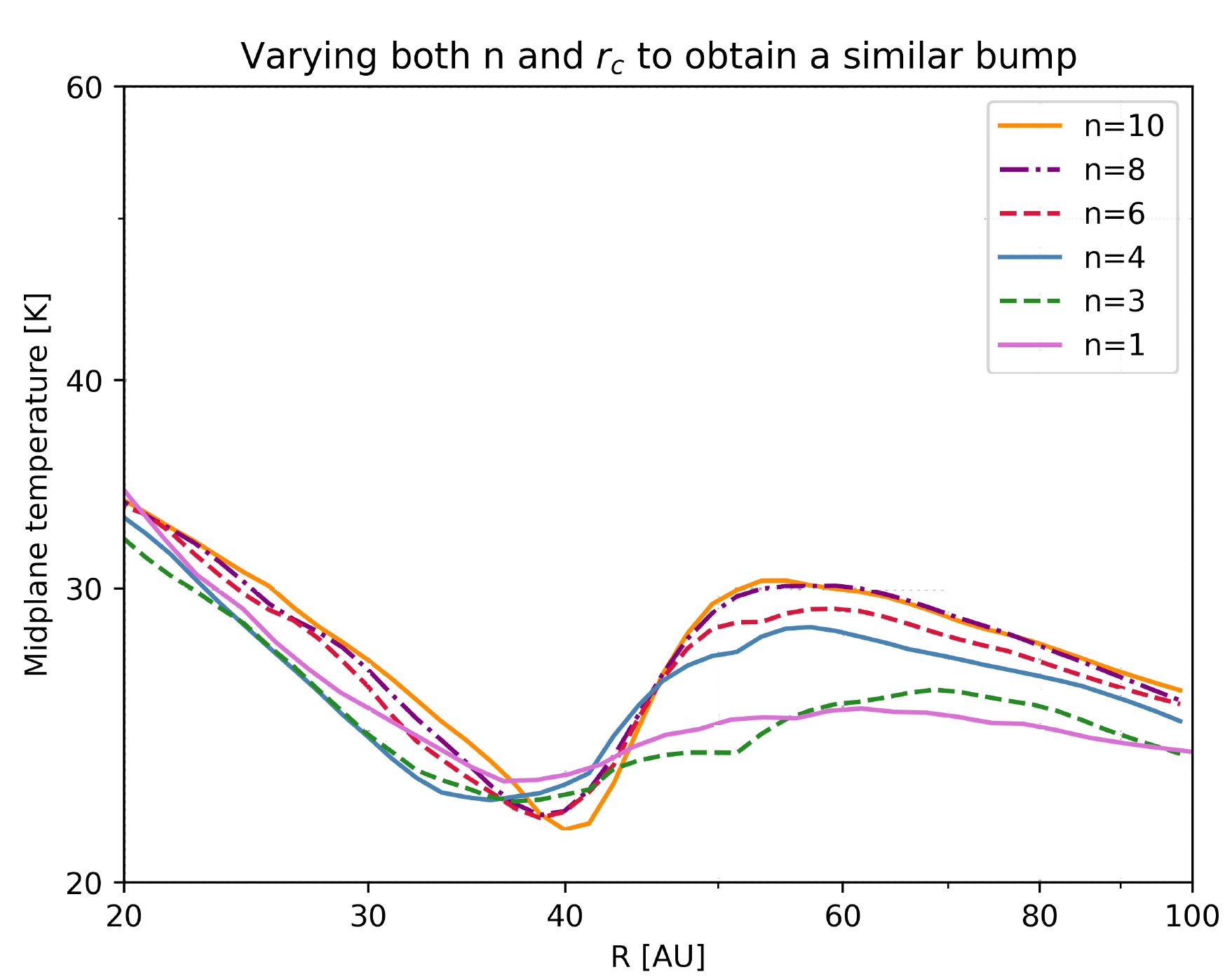}
    \caption{Temperature bump varying $n$ and $r_{\rm c}$ according to Eq.~\ref{eq:change_rc}. The bump position is the same for all cases, but the amplitude decreases for lower $n$.}
    \label{fig:bump_high_n}
\end{figure}

The main physical mechanism behind the temperature bump is the transition of the disk midplane from the optically thick to the optically thin regime with respect to stellar irradiation. In the optically thick region, the midplane is heated mainly by infrared photons reprocessed in the surface layers, whereas in the optically thin regime, direct stellar photons can reach deeper layers, leading to a local enhancement in the heating rate.
At zeroth order, the location of the bump can be estimated by finding the radius where the optical depth to stellar photons reaches $\tau_* = 2/3$. Approximating the optical depth as
\begin{equation}
    \tau_* \approx \frac{\kappa_* \Sigma(r)}{\sin\alpha},
\end{equation}
where $\kappa_*$ is the opacity at stellar wavelengths, $\Sigma(r)$ is the surface density, and $\alpha$ is the grazing angle \citep{Chiang_Goldreich}, here taken as a constant $\alpha = 0.02$ (its precise value has little effect on the result), one finds
\begin{equation}
    r_{\text{bump}}\approx r_{\rm c}\Bigg[\frac{1}{n}W\Bigg(n\Big(\frac{\kappa_*\Sigma_0 r_0}{r_{\rm c}\alpha}\Big)^n\Bigg)\bigg]^{1/n}
\end{equation}
where $W$ is the Lambert $W$-function \citep[see e.g.][]{Corless1996LambertW}. 
This expression reproduces the bump radius predicted by full radiative transfer (\texttt{RADMC-3D}) within $\approx 1-3$~AU. 
There is a degeneracy between $n$ and $r_{\rm c}$ in setting the bump location. For a fixed $n$, one can adjust $r_{\rm c}$ to align the $\tau_*=2/3$ radius across models:
\begin{equation}
    \frac{2}{3} = \frac{\kappa_* M_{\mathrm{disk}}\, n}{2\pi r_{\mathrm{bump}}\, r_{\rm c}^2 \sin\alpha}
    \exp\!\left[ -\left(\frac{r_{\mathrm{bump}}}{r_{\rm c}}\right)^n \right].
    \label{eq:change_rc}
\end{equation}
Figure \ref{fig:bump_high_n} shows \texttt{RADMC-3D} midplane temperatures for $n=1$--10, with $r_{\rm c}$ varied using Eq. \ref{eq:change_rc} so that all bumps occur at the same radius. The position matches by construction, but the amplitude of the bump decreases for lower $n$. This occurs because shallower profiles distribute mass more evenly across radii, producing a more gradual opacity transition and a broader, less pronounced $\tau_*$ crossing. As a result, the contrast in heating between inside and outside the bump location is smaller.

\begin{figure}
    \centering
    \includegraphics[width=\hsize]{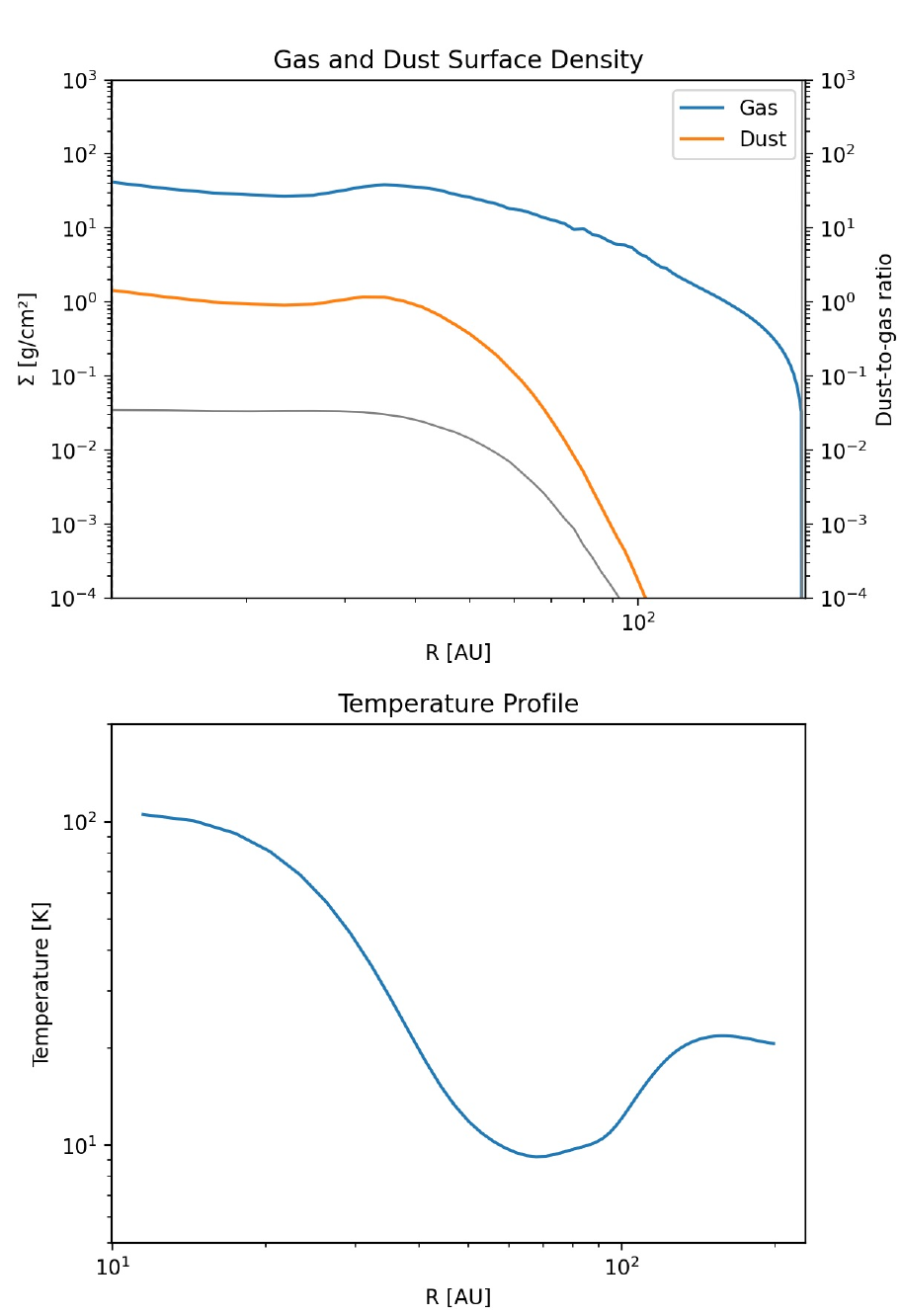}
    \caption{Top panel: gas and dust surface density profile for the $r_c=60$ AU simulation at 0.3 Myr. Bottom panel: gas midplane temperature profile for the $r_c=60$ AU simulation at 0.3 Myr.}
    \label{fig:rc60}
\end{figure}

The effect of varying $r_c$ while keeping $n$ fixed is to shift the location of the optical depth transition, and therefore the location of the bump itself: increasing $r_c$ moves the bump outward, since a larger characteristic radius shifts the surface density (and hence the $\tau_*=2/3$ crossing) further from the star. In this study we adopt $r_c=40$~AU, chosen to keep the bump as close to the star as possible -- thereby reducing computational cost, since the dynamical and thermal timescales at smaller radii are shorter -- while remaining sufficiently far from the inner boundary of the computational grid to avoid boundary effects (see Section~\ref{sec: hydro}).
As a proof of concept, we also ran a model with $r_c=60$~AU, keeping all other parameters unchanged. In this case, both the temperature bump and the corresponding dust density peak are located further out, consistent with the radial shift discussed above. The evolution also proceeds more slowly than in the fiducial $r_c=40$~AU case, as expected from the longer dynamical and growth timescales at larger orbital radii. Figure~\ref{fig:rc60} shows the simulation snapshot at $t=0.3$~Myr for this model: despite the slower evolution and the different bump location, the qualitative behavior is unchanged, with a dust ring forming at the location of the temperature bump, confirming that our results do not depend qualitatively on the specific choice of $r_c$.

\end{appendix}
\end{document}